\documentclass[twocolumn,superscriptaddress,english,prx,showpacs,longbibliography]{revtex4-2}

\usepackage[colorlinks=true,urlcolor=blue,citecolor=blue,linkcolor=blue]{hyperref}

\usepackage{graphicx}
\usepackage{dcolumn}
\usepackage{bm}
\usepackage{amsfonts}
\usepackage{xcolor}
\usepackage[normalem]{ulem}

\usepackage{amsmath}

\usepackage{booktabs} %
\usepackage{makecell} %
\usepackage{tikz}
\usetikzlibrary{positioning, arrows.meta, shapes.geometric, fit, backgrounds, calc}

\begin{document}

\title{Differentiable Maximum Likelihood Noise Estimation for Quantum Error Correction}

\author{Hanyan Cao}\thanks{These authors contributed equally to this work.}
\affiliation{Science, Mathematics and Technology Cluster, Singapore University
 of Technology and Design, 8 Somapah Road, 487372 Singapore}

\author{Dongyang Feng}\thanks{These authors contributed equally to this work.}
\affiliation{CAS Key Laboratory for Theoretical Physics, Institute of Theoretical Physics, Chinese Academy of Sciences, Beijing 100190, China.}

\author{Zixuan Lu}\thanks{These authors contributed equally to this work.}
\affiliation{School of Fundamental Physics and Mathematical Sciences, Hangzhou Institute for Advanced Study, UCAS, Hangzhou 310024, China}
\affiliation{CAS Key Laboratory for Theoretical Physics, Institute of Theoretical Physics, Chinese Academy of Sciences, Beijing 100190, China.}
\affiliation{School of Physical Sciences, University of Chinese Academy of Sciences, Beijing 100049, China}

\author{Cheng Ye}
\affiliation{CAS Key Laboratory for Theoretical Physics, Institute of Theoretical Physics, Chinese Academy of Sciences, Beijing 100190, China.}
\affiliation{School of Physical Sciences, University of Chinese Academy of Sciences, Beijing 100049, China}

\author{Feng Pan}\email{feng\_pan@sutd.edu.sg}
\affiliation{Science, Mathematics and Technology Cluster, Singapore University
 of Technology and Design, 8 Somapah Road, 487372 Singapore}


\begin{abstract}
Accurate noise estimation is essential for fault-tolerant quantum
computing, since decoding performance depends critically on the fidelity
of circuit-level noise parameters. We introduce a differentiable Maximum
Likelihood Estimation (dMLE) framework that learns these parameters
directly from experimental syndrome data using likelihood gradients, with no
decoder in the training loop. Mapping syndrome likelihoods to statistical-mechanical models enables exact and efficient evaluation via planar solvers for repetition codes and tensor networks (TNs) for surface codes, extending the exact treatment to code distance $d=5$ with a small number of rounds, and, via spatial subsampling combined with time-translation symmetry, to larger codes and larger round counts. Across four experimental superconducting
datasets, dMLE reduces logical error rates by up to $30.6\%$ for
repetition codes. On the Sycamore $d=3$ surface code, a TN decoder
using the dMLE prior attains a logical error rate comparable to that
reported for state-of-the-art AlphaQubit, using standard syndromes alone
without auxiliary leakage data. On the larger $d=5$ surface code the
dMLE prior lowers the logical error rate of the TN and Tesseract
decoders by $7.18\%$ and $5.23\%$ relative to a prior optimized by
reinforcement learning; at $d=7$ it lowers that of Tesseract by
$4.56\%$ and $4.83\%$ in the $X$ and $Z$ bases.

\end{abstract}

\maketitle

Fault-tolerant quantum computing (FTQC) promises computational capabilities far beyond classical reach.
Realizing this promise hinges on quantum error correction (QEC)~\cite{shor1995scheme, fowler2012surface, terhal2015quantum}, which protects fragile logical information by encoding it redundantly across many physical qubits and continuously measuring error syndromes~\cite{google2021exponential, google2023suppressing}.
However, QEC can only be as effective as the decoder that interprets these syndromes, and the decoder can only be as accurate as the noise model it assumes~\cite{dennis2002topological}.
In practice, the noise affecting a quantum processor is described by a detector error model (DEM) that assigns prior probabilities to each possible error mechanism in the circuit~\cite{gidney2021stim, mcewen2023relaxing}.
When these priors deviate from the true physical noise, decoding quality deteriorates sharply: Google's landmark surface-code experiment on the Willow processor demonstrated that accurately estimated circuit-level noise parameters were essential to achieving below-threshold error suppression~\cite{google2024belowThreshold}, and a dedicated study of decoder priors has shown that even modest prior inaccuracies can erode the exponential suppression expected from increasing code distance~\cite{RL-PhysRevLett.133.150603}.
Accurately estimating DEM priors from experimental data is therefore the critical bottleneck standing between today's noisy hardware and the full error-suppression power of QEC.

The challenge of noise estimation lies in the inverse nature of the problem: physical errors are not directly observable; only their degenerate mappings, syndromes, are measured. 
Consequently, the task is to infer the underlying error probabilities that maximize the likelihood of the observed syndrome statistics. 
While this is fundamentally a parameter estimation problem, the high-dimensional space of error configurations makes direct inference computationally intractable for general codes.

Previous efforts primarily rely on correlation analysis (CA) and reinforcement learning (RL). Correlation-based methods efficiently estimate error rates from local syndrome statistics~\cite{spitz2018adaptive, google2021exponential, google2023suppressing, remm2025experimentally, blume2025estimating,takou2025estimating,arms2025estimating,bhardwaj2025adaptive}. However, these approaches do not optimize a global likelihood over the full joint distribution; most of them are restricted to pairwise interactions, and thereby struggle with codes characterized by higher stabilizer weights, such as color codes~\cite{lacroix2025scaling}. They are also prone to generating unphysical negative probabilities~\cite{RL-PhysRevLett.133.150603}. Alternatively, RL optimizes decoder priors using the logical error rate as a reward~\cite{RL-PhysRevLett.133.150603,sivak2025reinforcement}, which means it necessitates running a fast, often sub-optimal decoder in its training loop. While adaptable, RL parameters lack clear correspondence to physical error mechanisms and critically depend on a high-quality initialization such as CA~\cite{RL-PhysRevLett.133.150603}. Moreover, our benchmarks show that the gains obtained with RL priors depend on the decoder: priors optimized with matching-based decoders perform less well than the dMLE prior with several other decoding algorithms. This is consistent with decoder-specific tuning, but does not by itself establish overfitting.

\begin{figure*}[htbp!]
    \centering
    \includegraphics[width=0.95\linewidth]{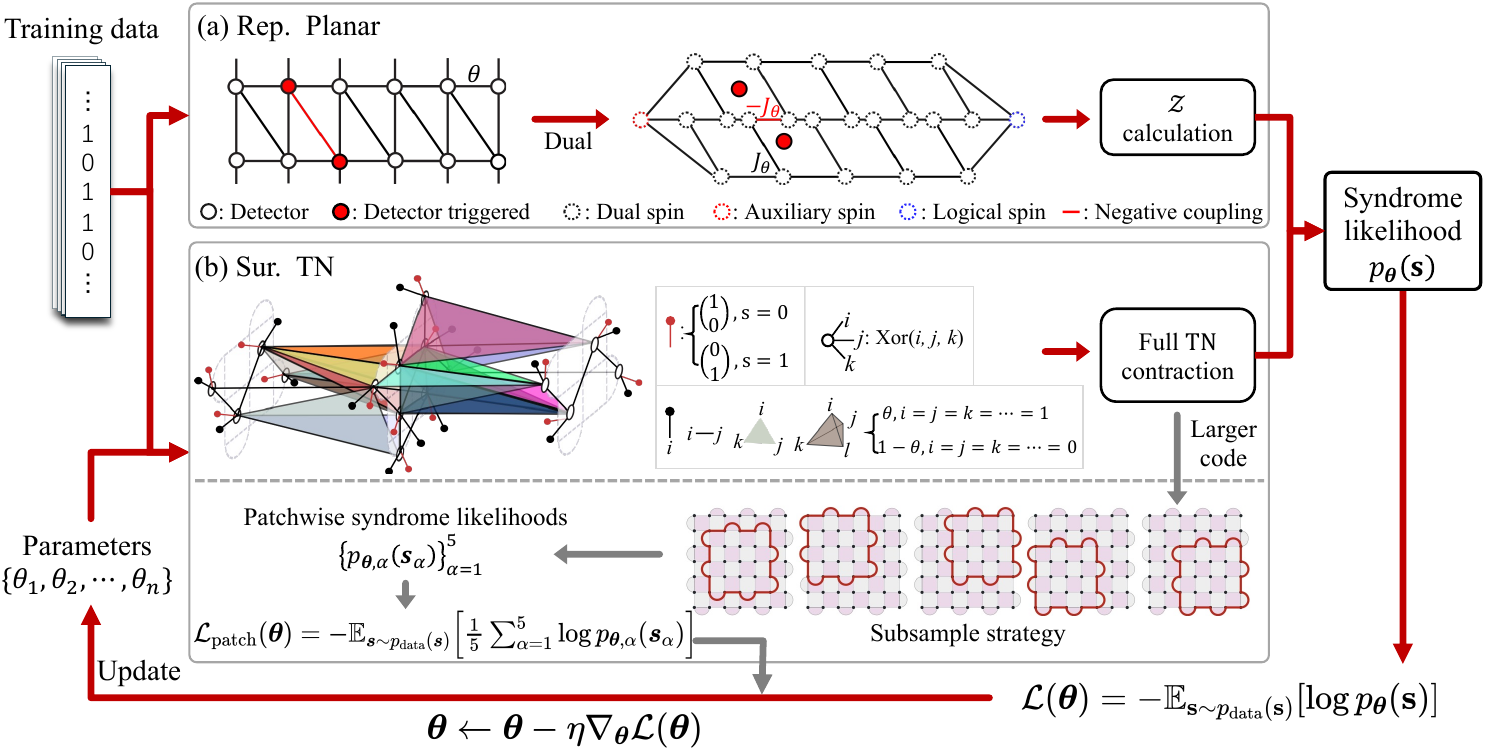}
    \caption{\textbf{Differentiable maximum-likelihood noise estimation (dMLE) framework.} Observed detection events $\mathbf{s}$ and DEM parameters $\bm{\theta}$ are mapped to either a dual-spin model for repetition codes (a) or a Tensor Network (TN) for surface codes (b). In (a), the likelihood is obtained from the partition function $\mathcal{Z}$. In the upper part of (b), a full TN contraction evaluates the surface code syndrome likelihood $p_{\bm{\theta}}(\mathbf{s})$; in the lower part, a $d=7$ code is spatially decomposed into five overlapping $d=5$ patches, whose exact marginal likelihoods define the composite loss $\mathcal{L}_{\mathrm{patch}}(\bm{\theta})$. Full-network training uses the likelihood gradient; patch-based training uses masked gradient contributions from patches fully containing each mechanism (see Supplemental Material).}
    \label{fig:1}
\end{figure*}

In this Letter, we propose a physics-inspired framework that estimates
the DEM prior by differentiable Maximum Likelihood Estimation (dMLE).
Mapping syndrome likelihoods onto partition functions of statistical-mechanical spin models lets us evaluate them exactly for repetition codes, via the \textit{Planar} solver~\cite{cao2025exact}, and for small surface codes, via tensor networks (TNs), rendering the physical error probabilities directly optimizable. Spatial subsampling and time-translation symmetry then extend the treatment approximately to larger distances and more rounds. Unlike correlation-based or RL-based
methods, dMLE uses likelihood gradients of the observed syndromes to learn
parameters within the physical domain, without a decoder in the training loop.
Full-network training minimizes the negative log-likelihood; at $d=7$, we use
the masked patch-based update described below.
We benchmark the resulting prior, hereafter the dMLE prior, on
experimental repetition-code data and on Google's Sycamore
$d=3$~\cite{dataset_old} and $d=5$~\cite{dataset} and Willow
$d=7$~\cite{GoogleQuantumAI2024WillowData} surface-code data. It
transfers across decoders without retraining and outperforms the RL and
correlation-based priors for most of the decoders tested; on the $d=3$ code it brings a
TN decoder that uses standard syndromes alone to a logical error rate
comparable to that reported for AlphaQubit~\cite{bausch2024learning}.

\textit{Maximum Likelihood Noise Estimation.---}
In circuit-level noise, conventional Pauli error models on qubits often fail to
capture the dynamics of fault-tolerant protocols. We instead use the DEM,
which characterizes the probabilistic relations between error
mechanisms and detectors (syndrome combinations) within a spacetime
circuit~\cite{mcewen2023relaxing}. A DEM is a hypergraph: nodes are detectors,
and hyperedges are independent error mechanisms $e_i$ with probabilities
$\theta_i$. Decomposing the noise into $n$ independent mechanisms, an error configuration is
$\mathbf{e}\in\{0,1\}^n$, and our goal is to estimate the priors
$\bm{\theta}=\{\theta_i\}$ from observed syndromes $\{\mathbf{s}^{(k)}\}$. The
probability of a syndrome $\mathbf{s}$ is the sum over all consistent error
configurations,
\begin{equation}\label{eq:sum}
p_{\bm{\theta}}(\mathbf{s}) = \sum_{\mathbf{e} \in \mathcal{C}(\mathbf{s})} p_{\bm{\theta}}(\mathbf{e}),
\end{equation}
where $\mathcal{C}(\mathbf{s})$ is the set of errors causing $\mathbf{s}$. With
error probabilities identified as Boltzmann weights, this sum equals the
partition function of a statistical-mechanical spin
model~\cite{chubb2021statistical}. We minimize the negative log-likelihood,
\begin{equation}
\mathcal{L}(\bm{\theta}) = -\mathbb{E}_{\mathbf{s}\sim p_{\text{data}}(\mathbf{s})}[\log p_{\bm{\theta}}(\mathbf{s})],
\end{equation}
by gradient descent.

The key challenge is evaluating Eq.~\eqref{eq:sum}, which is in general
$\#$P-hard~\cite{cao2025exact,chubb2021statistical,PhysRevX.2.021004}. For
planar DEMs of repetition codes, the \textit{Planar}
method~\cite{cao2025exact} evaluates it exactly in polynomial time via a dual
spin model [Fig.~\ref{fig:1}(a)]. For surface codes, where \textit{Planar} does
not apply, we use tensor networks (TNs), whose product-sum
contraction~\cite{PhysRevA.90.032326,chubb2021generaltensornetworkdecoding,PRXQuantum.5.040303,google2023suppressing}
yields the exact $p_{\bm{\theta}}(\mathbf{s})$ [upper part of Fig.~\ref{fig:1}(b)]:
each detector is an XOR tensor enforcing even parity, red dangling edges impose
detection-event constraints, and probability tensors assign weights $\theta_i$ and
$1-\theta_i$. Both \textit{Planar} and TN summations are fully differentiable in $\bm{\theta}$,
enabling end-to-end gradient refinement toward the maximum-likelihood estimate.

As the code distance increases, a single exact TN contraction becomes
intractable. We therefore decompose the larger code spatially into
smaller overlapping $d=5$ patches with substantially reduced
contraction complexity, all sharing a single physical parameter vector
$\bm{\theta}$ [lower part of Fig.~\ref{fig:1}(b)]. A TN contraction
for each patch gives the likelihood of its local syndrome. And $\bm{\theta}$ is updated using masked likelihood gradients: each
parameter receives contributions only from patches fully containing its
mechanism, summed and divided by the total number of patches $N_p$. The patch size is
chosen as the smallest one guaranteeing full coverage: every error
mechanism in the larger code's DEM is fully contained in at least one
$d=5$ patch, whereas $d=3$ patches are insufficient for this coverage condition.
Coverage ensures that each parameter has an unmasked patch contribution; it
does not by itself establish identifiability from the patch marginals. The patch
likelihoods and their gradient contributions can be evaluated
independently and parallelized across GPUs; further details and a
complexity analysis are provided in the Supplemental
Material~\cite{sm}.


\textit{Numerical Results.---}
\begin{figure}
    \centering
    \includegraphics[width=0.85\linewidth]{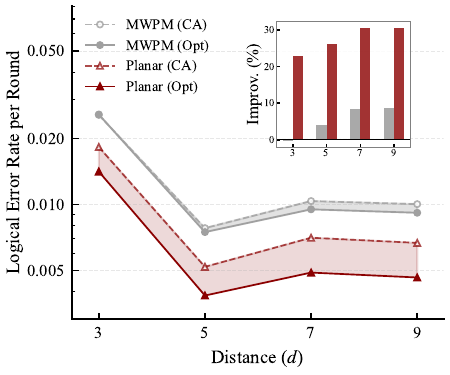}
     \caption{\textbf{Validation on BAQIS superconducting repetition codes.} Logical error rate per round versus code distance $d\in\{3,5,7,9\}$. Dashed lines (CA) correspond to the initial prior derived from correlation analysis~\cite{google2021exponential}, while solid lines (Opt) represent the dMLE prior. Inset: relative reduction of the logical error rate per round from the CA prior to the dMLE prior, for MWPM (gray bars) and \textit{Planar} (red bars).}
    \label{fig:rep}
\end{figure}
To test dMLE under realistic experimental noise, we first use superconducting repetition-code data from BAQIS~\cite{cao2025exact}. Figure~\ref{fig:rep} shows the logical error rate per round versus code distance. The rate is not monotonic in $d$: longer chains include additional lower-quality qubits, and errors beyond the independent Pauli model become more prominent as the system grows~\cite{cao2025exact}. We therefore assess the effect of prior optimization at each distance. The dMLE prior reduces the logical error rate of the \textit{Planar} decoder at every distance and that of Minimum Weight Perfect Matching (MWPM)~\cite{higgott2021pymatching,higgott2023Sparse} for $d\geq5$, leaving MWPM essentially unchanged at $d=3$.

The gains are more pronounced for \textit{Planar}, as illustrated by the separation between the initial and optimized curves and quantified in the inset. MWPM improves by up to $8.66\%$, whereas the reduction for \textit{Planar} reaches $30.6\%$ at larger distances. These results show that likelihood-based optimization extracts noise information from experimental syndromes that improves error suppression, with the size of the benefit depending on how the decoder uses the estimated prior.

\begin{figure*}[t!]
 \includegraphics[width=0.9\linewidth]{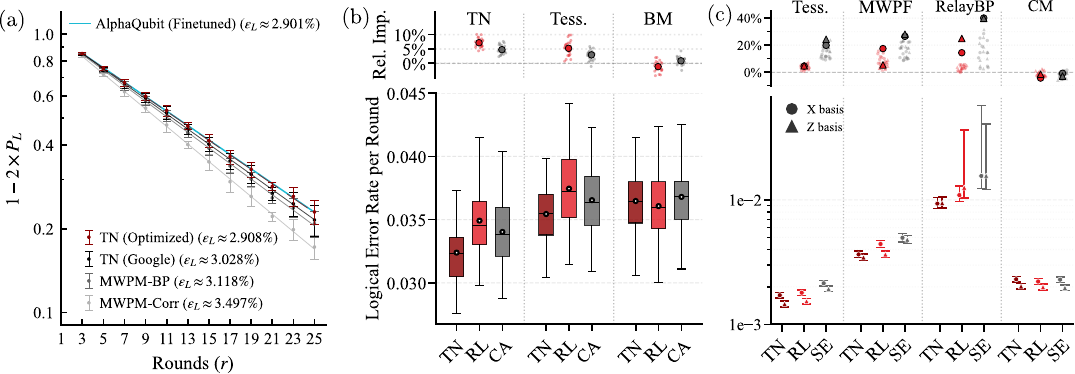}
 \caption{\textbf{Numerical results on surface-code experimental datasets.} (a) Decoding performance on Google's Sycamore $d=3$ surface-code dataset~\cite{dataset_old}. The logical fidelity $1-2P_L$, with $P_L$ the logical error probability, is plotted versus the number of syndrome rounds $r$; the logical error rate per round $\epsilon_L$ is obtained from exponential fits. The results for Google's TN decoder [TN (Google)], belief matching (BM, labelled MWPM-BP), and correlated matching (CM, labelled MWPM-Corr) are obtained directly from the dataset. The AlphaQubit line is the exponential decay reconstructed from the $\epsilon_L$ reported in Ref.~\cite{bausch2024learning}. The TN (Optimized) result (red) is obtained with an exact TN decoder using the dMLE prior. (b) Google's Sycamore $d=5$ dataset used for the RL method~\cite{dataset}. Lower panel: $\epsilon_L$ for the TN, Tesseract (Tess.), and BM decoders, each using three priors (horizontal axis): the dMLE prior (labelled TN), the RL prior trained with BM (RL), and the CA prior (CA). The RL and CA priors and the BM results are taken from the dataset. Each box summarizes the 24 subsets used. Upper panel: relative improvement $(\epsilon_L^{\mathrm{ref}}-\epsilon_L^{\mathrm{dMLE}})/\epsilon_L^{\mathrm{ref}}$ of the dMLE prior over the reference prior, RL (red) or CA (gray); light markers show individual subsets and solid markers their average. (c) Google's Willow $d=7$ dataset~\cite{GoogleQuantumAI2024WillowData}, decoded with Tesseract, Minimum-Weight Parity Factor (MWPF)~\cite{mwpf}, RelayBP~\cite{relaybp}, and CM. Lower panel: $\epsilon_L$ for the three priors---the dMLE prior, the RL prior, and the syndrome-estimated (SE) prior of Ref.~\cite{takou2026logical}---in the $X$ and $Z$ bases. Upper panel: relative improvement of the dMLE prior over the reference prior, RL or SE; light markers show the improvement of $P_L$ at each individual number of rounds, and solid markers that of the fitted $\epsilon_L$. Error bars in the lower panel indicate 95\% confidence intervals. The RL prior is taken from the dataset; SE extends pairwise CA to multi-detector correlations~\cite{takou2026logical}.}
 \label{fig:data}
\end{figure*}

We next turn to Google's Sycamore $d=3$ surface-code dataset~\cite{dataset_old}. As shown in Fig.~\ref{fig:data}(a), the TN decoder with the dMLE prior achieves a logical error rate per round of $\epsilon_L=2.908\%$, compared with $3.028\%$ for Google's TN decoding of the same data and $2.901\%$ reported for AlphaQubit~\cite{bausch2024learning}. Our DEM estimation uses standard two-state syndrome data alone, whereas AlphaQubit's training additionally uses auxiliary experimental information, including leakage-resolving three-state readout. The improvement over Google's TN result comes from optimizing the DEM prior against the observed syndrome distribution.

Beyond direct decoding, the learned DEM provides an experimentally aligned generative model that could supply samples for pre-training neural decoders, potentially improving their data efficiency and generalization~\cite{bausch2024learning}. Neural decoders can also accept richer inputs. Extending the generative model to include auxiliary information of the kind used by AlphaQubit, such as leakage-resolving readout, is therefore a natural next step that may further improve decoding performance.

We next consider the Sycamore $d=5$ dataset used in Google's RL study~\cite{dataset}, with $r=5$--$25$. Its experimental settings differ from those of the $d=3$ dataset~\cite{google2023suppressing}. Walsh--Hadamard decompositions of XOR tensors simplify the TN to Hadamard matrices and probability vectors~\cite{PRXQuantum.5.040303}. Optimized contraction orders~\cite{pan2022Simulation,pan2022Solving,pan2024Efficient,kalachev2022multitensorcontractionxebverification}, reused across syndromes and parameter updates, enable uncompressed GPU-batched likelihood evaluation. Reverse-mode differentiation computes gradients for the full DEM parameter vector in one backward pass~\cite{liao2019differentiable}. These features make repeated likelihood-based fitting practical; details are given in the Supplemental Material~\cite{sm}. We train on $r=5$ and transfer the prior to larger round numbers using approximate time-translation symmetry~\cite{Kelly_2015,RL-PhysRevLett.133.150603}.

Within our computational budget, we evaluate 24 of the dataset's 53 subsets, initializing from the supplied uniform depolarizing DEM (\texttt{simple.dem}, error strength $10^{-3}$)~\cite{RL-PhysRevLett.133.150603}. Figure~\ref{fig:data}(b) compares TN, Tesseract~\cite{beni2025tesseractdecoder}, and belief matching (BM)~\cite{belief_matching} using dMLE, RL trained with BM, and CA priors. Relative to RL and CA, the mean reductions in $\epsilon_L$ are $7.18\%$ and $4.78\%$ for TN, and $5.23\%$ and $2.94\%$ for Tesseract. BM improves only marginally over CA and does not improve over RL. Thus the same dMLE prior transfers across decoders without retraining, although the gains depend on the decoder.

Finally, we apply the masked update to Google's Willow $d=7$ dataset~\cite{GoogleQuantumAI2024WillowData}, using five overlapping $d=5$ patches because full TN contraction is out of reach [Fig.~\ref{fig:1}(b)]. We train separately for the $X$ and $Z$ bases on 50,000 shots at $r=10$. Figure~\ref{fig:data}(c) compares Tesseract, MWPF, RelayBP, and CM using the dMLE prior, Google's RL prior optimized with CM, and the syndrome-estimated (SE) prior of Ref.~\cite{takou2026logical}. We evaluate $r=13$--$250$ and fit $\epsilon_L$ for each decoder, prior, and basis, restricting RelayBP to $r\leq190$.

dMLE lowers the fitted $\epsilon_L$ in both bases for all three non-matching decoders. Relative to RL, the reductions are $4.56\%$ and $4.83\%$ for Tesseract in the $X$ and $Z$ bases and reach $25.0\%$ for RelayBP; relative to SE, they range from $19.9\%$ to $40.0\%$. For MWPF in the $Z$ basis, the reductions relative to RL and SE are $5.46\%$ and $28.05\%$. CM is the exception: RL and SE yield slightly lower $\epsilon_L$ than dMLE; the dMLE rate is $0.730\%$ to $4.27\%$ higher. The RL advantage with CM may partly reflect its direct use of CM during training~\cite{GoogleQuantumAI2024WillowData}. SE likewise performs slightly better only with the matching-based decoder tested, while performing worse with Tesseract, MWPF, and RelayBP, indicating more limited transferability across these decoders.

For all three surface-code datasets, training from the dataset-supplied and correlation-based priors (CA at $d=3,5$ and SE at $d=7$) yields consistent results, providing empirical evidence of convergence consistency for the two initializations tested~\cite{sm}.

\textit{Conclusion.---}
In this work, we introduced dMLE, a maximum-likelihood framework for estimating circuit-level noise directly from experimental syndrome data. Exact syndrome likelihoods are tractable for repetition codes and surface codes up to $d=5$ with a limited number of rounds. Beyond this regime, masked spatial subsampling combined with approximate time-translation symmetry extends the framework to surface-code memory experiments on current hundred-qubit devices. Differentiation with respect to physical error parameters yields physically interpretable priors, and training requires no decoder.

Benchmarks on four experimental superconducting datasets, from repetition codes to the Willow $d=7$ surface code, demonstrate improvements across decoders with distinct inference mechanisms using the same learned prior without retraining. The gains vary, and matching-based decoders do not consistently improve over optimized baselines. These results support the practical transferability of priors learned without decoder-specific training. On Sycamore $d=3$, the dMLE prior allows a TN decoder using standard syndromes alone to reach a logical error rate comparable to that reported for AlphaQubit, while spatial subsampling reaches code distances beyond exact contraction. The matching-based exceptions motivate further study of interactions between prior estimation and the decoder's treatment of the DEM.

The spatial structure and approximately repeated portions of lattice-surgery circuits also motivate extending patch-based estimation beyond memory experiments; adapting patches to changing boundaries and operations remains future work. Beyond decoding, accurate, physically grounded priors can provide experimentally aligned training data for neural decoders~\cite{bausch2024learning,cao2025generative,battistel2024using,senior2025scalable,zhang2026learningdecodeparallelselfcoordinating}. The differentiability of the likelihood also offers a route to gate- and pulse-level noise estimation~\cite{sivak2025reinforcement}, potentially enabling joint optimization of physical controls and error-correction performance. We plan to explore these directions in future work.

\textit{Acknowledgements.---}
The authors would like to thank Haifeng Yu and Huikai Xu for providing access to their experimental data, and Pan Zhang and Zhiheng Zhang for helpful discussions.
This work is supported by the Ministry of Education Singapore under grant No. SKI 2021\_07\_03,  National Quantum Computing Hub translational fund No. W24Q3D0002 and MTC Young Individual Research Grant No. H25-MRG3466.

The Python implementation of our algorithm is available at~\cite{code}.

\bibliography{main}%

\clearpage

\begin{center}
	\textbf{\large Supplemental Material for ``Differentiable Maximum Likelihood Noise Estimation for Quantum Error Correction''}
\end{center}
\bigskip
\twocolumngrid

\section{Methods}

\subsection{Statistical-Mechanical Mapping and Numerical Validation}
\label{sec:statmech}
In this section, we elucidate the relationship between the syndrome probability $p_{\bm{\theta}}(\mathbf{s})$ and the partition function $\mathcal{Z}$ of the corresponding spin glass model. The spin coupling is defined as:
\begin{equation}
    J_i = (-1)^{e_i} \frac{1}{2}\log{\frac{1-\theta_i}{\theta_i}} \;,
\end{equation}
where $e_i$ denotes the $i$th component of a length-$n$ error configuration $\mathbf{e}\in\{0, 1\}^n$, and $\theta_i$ represents the probability of this error component being flipped. The corresponding Hamiltonian for a given $\mathbf{e}$ is expressed as:
\begin{equation}\label{eq:h}
    H_{\mathbf{e}} (\tilde{\mathbf{s}}) = - \sum^n_{i=1} J_i \overset{\leftharpoonup}{\tilde{s}}_i \overset{\rightharpoonup}{\tilde{s}}_i - \frac{1}{2}\log{\prod^n_{i=1} \theta_i(1-\theta_i)} \;,
\end{equation}
where $\tilde{\mathbf{s}}\in\{\pm1\}^{n-m+1}$ is the spin configuration, $\overset{\leftharpoonup}{\tilde{s}}_i$ and $\overset{\rightharpoonup}{\tilde{s}}_i$ are two spins adjacent to edge $i$. The second term serves as a spin-independent constant. We can then derive the probability of a specific error configuration $\mathbf{e}$ as follows:
\begin{equation}
\begin{split}
    &\mathrm{exp}\{{-H_{\mathbf{e}}(\tilde{\mathbf{s}}=\mathbf{1})}\} \\
    &=\exp \{ \sum^n_{i=1} (-1)^{e_i} \frac{1}{2}\log{\frac{1-\theta_i}{\theta_i}}\} \cdot \prod^n_{i=1}[\theta_i(1-\theta_i)]^\frac{1}{2}\\
    &=\exp \{\frac{1}{2} \log\prod^n_{i=1} \left(\frac{1-\theta_i}{\theta_i}\right)^{1-2e_i}\}\cdot \prod^n_{i=1}[\theta_i(1-\theta_i)]^\frac{1}{2}\\
    &=\prod^n_{i=1}\left(\frac{1-\theta_i}{\theta_i}\right)^{\frac{1}{2}-e_i}\cdot [\theta_i(1-\theta_i)]^\frac{1}{2}\\
    &=\prod^n_{i=1} \theta^{e_i}_i\cdot(1-\theta_i)^{1-e_i}\\
    &=p_{\bm{\theta}} (\mathbf{e})\;.
\end{split}
\end{equation}

\begin{figure}[htbp!]
    \centering
    \includegraphics[width=0.95\linewidth]{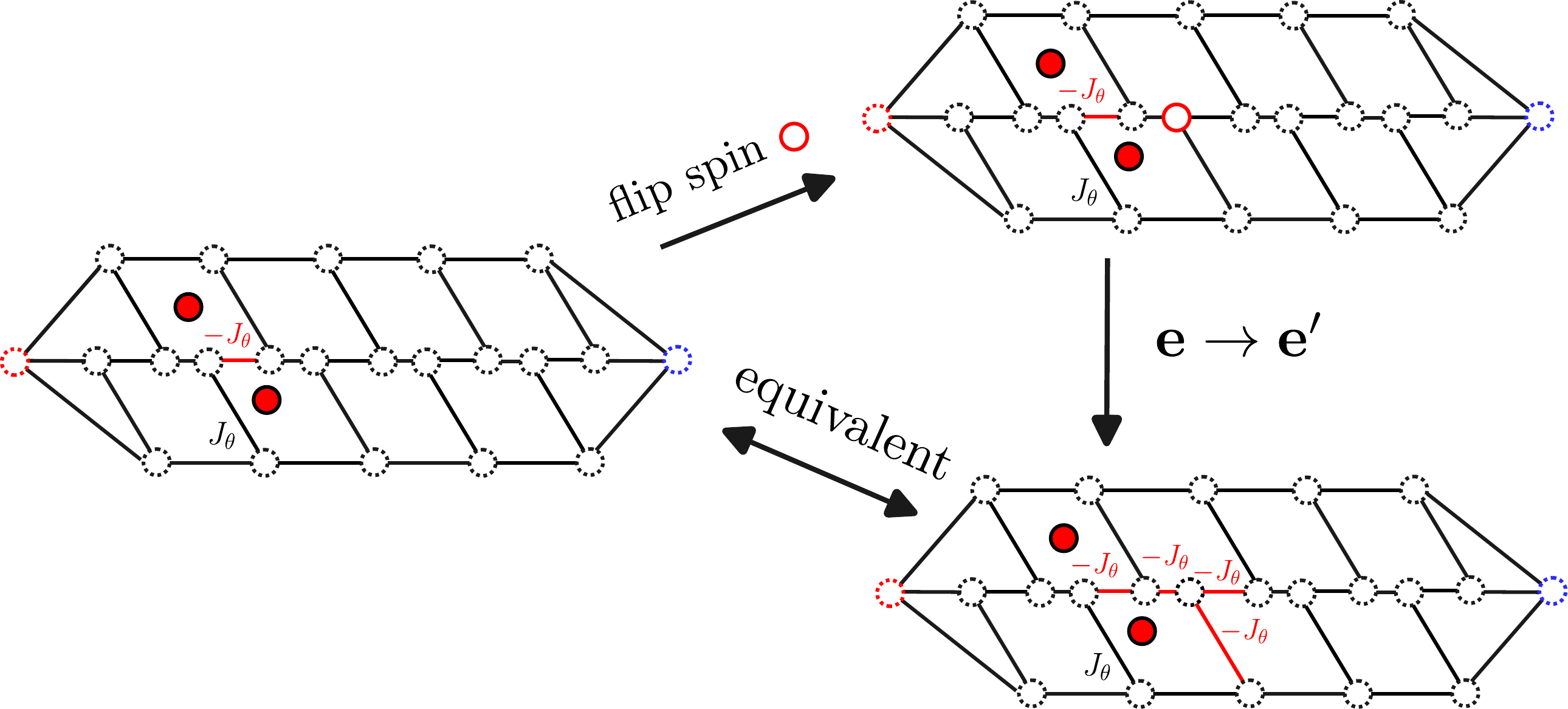}
    \caption{Equivalent transformation of Ising model.}
    \label{fig:s1}
\end{figure}

Consider the scenario where a spin is flipped, as illustrated in Fig.~\ref{fig:s1}; this operation changes the signs of the corresponding terms in the Hamiltonian Eq.~(\ref{eq:h}). This transformation is equivalent to redefining the Hamiltonian using a new error configuration $\mathbf{e}'$, which is obtained by applying an operator corresponding to the flipped spin onto the original error configuration $\mathbf{e}$.

One can verify that this new error configuration will also trigger the same set of detectors (e.g., the two red detectors in Fig.~\ref{fig:s1}). This conclusion holds generally for any sequence of spin flips. Consequently, the total probability of the observed detection events is one half of the partition function of this Ising model:
\begin{equation}
    p_{\bm{\theta}}(\mathbf{s}) = \frac{1}{2} \sum_{\tilde{\mathbf{s}}} \exp{\{-H_{\mathbf{e(\mathbf{s})}}(\tilde{\mathbf{s}})\}}\;,
\end{equation}
where $\mathbf{e}(\mathbf{s})$ is a reference error configuration consistent with $\mathbf{s}$. The factor $\frac{1}{2}$ arises because the introduction of the auxiliary spins endows the system with a $Z_2$ symmetry, which leads to the double-counting of all possible error configurations. As a result, the Kac-Ward formula can be employed to calculate this partition function exactly.

\begin{figure}[h]
    \centering
    \includegraphics[width=0.9\linewidth]{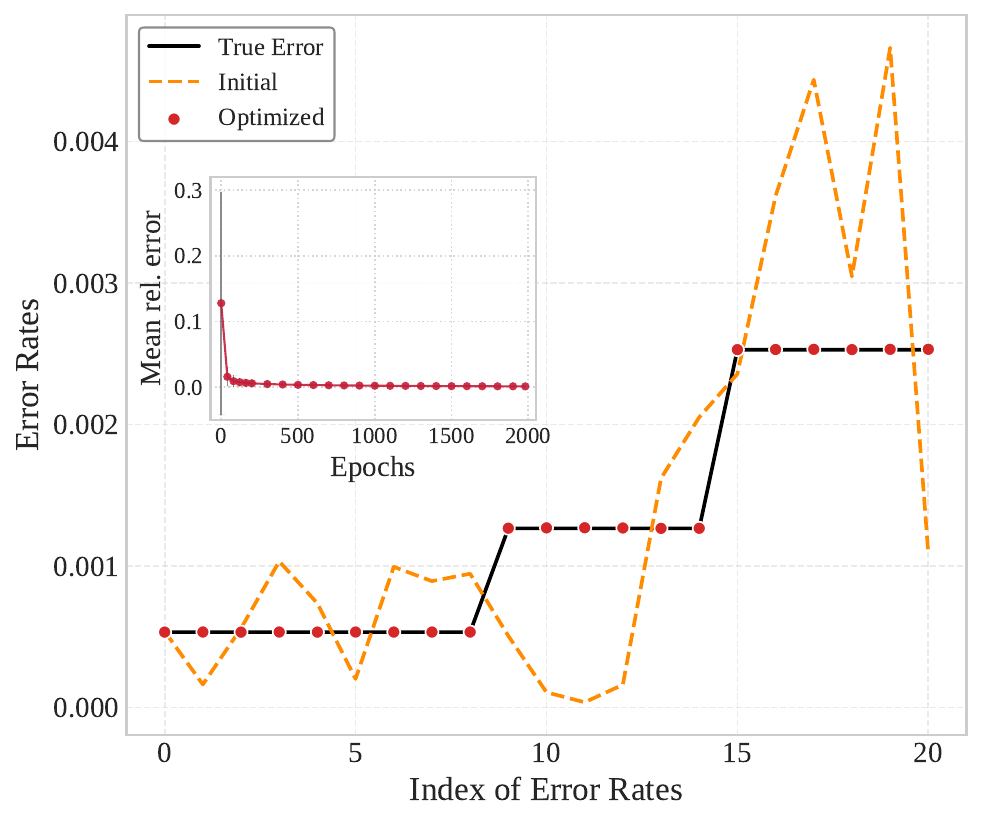}
    \caption{Benchmark on a $d=3$, $r=3$ repetition code. This model contains $8$ detectors and $256$ detector configurations. For every configuration $\mathbf{s}$, we employed the \textit{Planar} algorithm to exactly compute the probabilities $p_{\mathrm{data}}(\mathbf{s})$ and $p_{\bm{\theta}}(\mathbf{s})$, yielding an exact NLL loss. The orange line represents the initial random perturbative values. The red dots and black line represent converged results and the ground truth values. Inset: mean relative error of the parameters over the 2000 training epochs.}
    \label{fig:toy}
\end{figure}

Here we use a simple model to validate the ability of our gradient-based optimization to recover the underlying prior parameters from simulation data using a $d=3$ and $r=3$ repetition code under depolarizing circuit-level noise with error rate $\epsilon_p=0.001$. This benchmark model, generated from Stim, comprises $8$ detectors and $2^8=256$ detector configurations. The manageable system size allows for the direct computation of the complete probability table $p_{\mathrm{data}}({\mathbf{s}})$, yielding the exact negative log-likelihood (NLL) loss function. Consequently, we can optimize the parameters $\bm{\theta}$ to recover the prior probabilities with near-exact precision, as shown in Fig.~\ref{fig:toy}. The parameters were initialized as indicated by the orange dashed line. We then calculated $p_{\mathrm{data}}(\mathbf{s})$ for each detector configuration $\mathbf{s}$ using the true prior probabilities and iteratively updated $\bm{\theta}$ until the loss function converged. Finally, the parameters converged to the red dots, which exhibit excellent agreement with the true prior probabilities (black line). The inset illustrates the training dynamics, showing the monotonic decrease of the average relative error between the optimized and true priors until convergence. This result demonstrates that, for small-scale models, gradient descent effectively optimizes parameters to near-exact values. 

Subsequently, we extended our numerical evaluations to larger models, for which exact gradient computation is intractable.
\begin{figure}[htbp]
    \centering
    \includegraphics[width=0.8\linewidth]{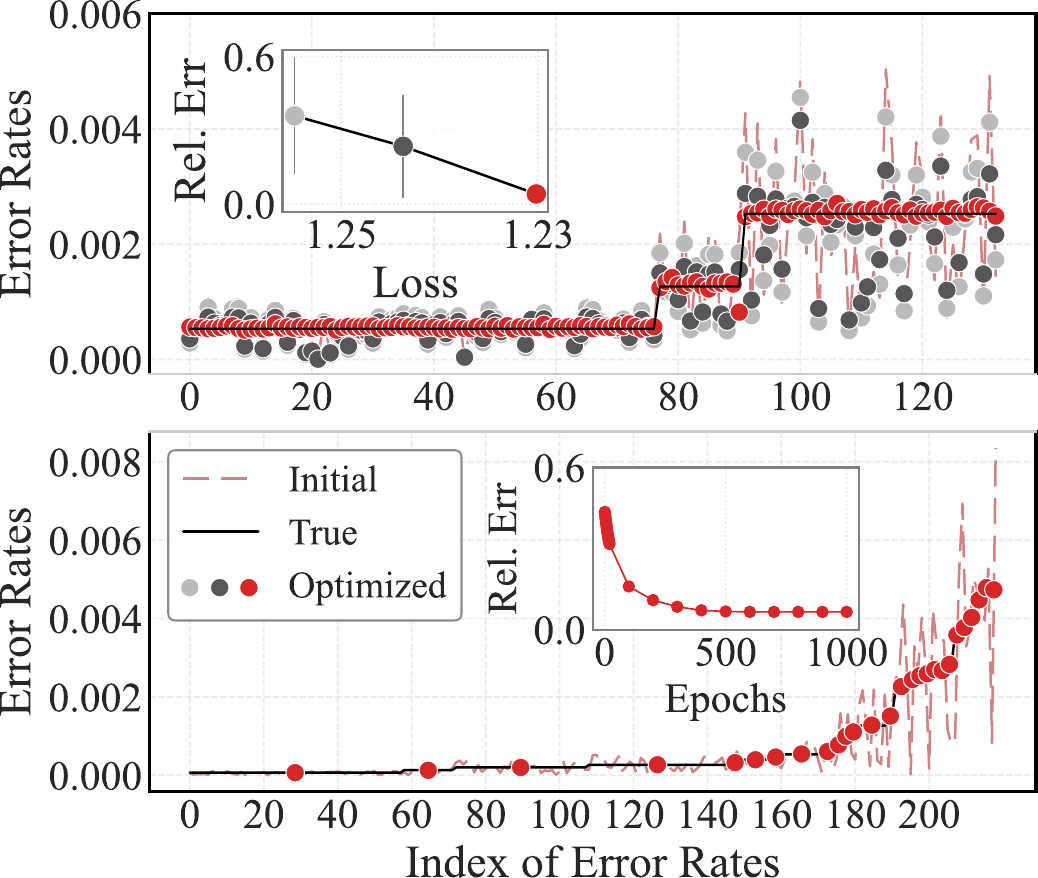}
    \caption{ (Upper) Optimization dynamics for a $d=7, r=7$ repetition code. The noisy initialization (dashed line) evolves during training (colored dots, from gray to red), with the final optimized parameters (red dots) closely matching the ground truth (black lines). The inset illustrates the positive correlation between the optimization objective (negative log-likelihood, NLL) and parameter accuracy (relative error). (Lower) Surface code DEM reconstruction ($d=3, r=3$, base depolarizing error $\epsilon_p=0.001$) using tensor-network (TN) contraction. To reduce visual clutter, the optimized parameters (red dots) display the mean values grouped by their corresponding ground-truth error rates.}
    \label{fig:sim}
\end{figure}
We first tested our method via numerical simulations on a $d=7, r=7$ repetition code, employing standard Monte Carlo gradients to approximately optimize the loss function. As shown in Fig.~\ref{fig:sim} (upper), despite starting from a highly perturbed initialization, the estimated priors progressively and accurately converge toward the ground truth. The inset further validates our core premise: minimizing the approximate NLL directly drives a corresponding reduction in the relative error of the estimated parameters.

Next, we evaluated our optimization scheme on surface codes. Because the complex structure of surface code DEMs renders the exact \textit{Planar} method inapplicable, we utilized TNs (Sec.~\ref{sec:tn}) to compute the syndrome likelihood. We simulated a $d=3, r=3$ rotated surface code (Z-basis memory experiment, $\epsilon_p=0.001$) using Stim~\cite{gidney2021stim}. Applying an identical random perturbation scheme to the initial priors, the optimized parameters rapidly and tightly converge to the analytical optima, as shown in Fig.~\ref{fig:sim} (lower), demonstrating the robustness and effectiveness of our approach even within the highly complex error structures of surface codes.

\subsection{Tensor Network}
\label{sec:tn}
The approach of Sec.~\ref{sec:statmech}, grounded in the mapping to statistical physics, fundamentally entails identifying a reference error configuration consistent with $\mathbf{s}$, applying operators that leave $\mathbf{s}$ invariant, and summing over this set of operators. This framework is referred to in the literature as the generator picture~\cite{PRXQuantum.5.040303}. In the context of circuit-level noise, although it is theoretically possible to formulate a tensor network within the generator picture, a significant challenge arises: for complex systems such as the surface code, constructing dual variables for the detectors while maintaining a sufficiently low average degree—a prerequisite for efficient contraction—remains a non-trivial task.
\begin{figure}[h]
    \centering
    \includegraphics[width=0.8\linewidth]{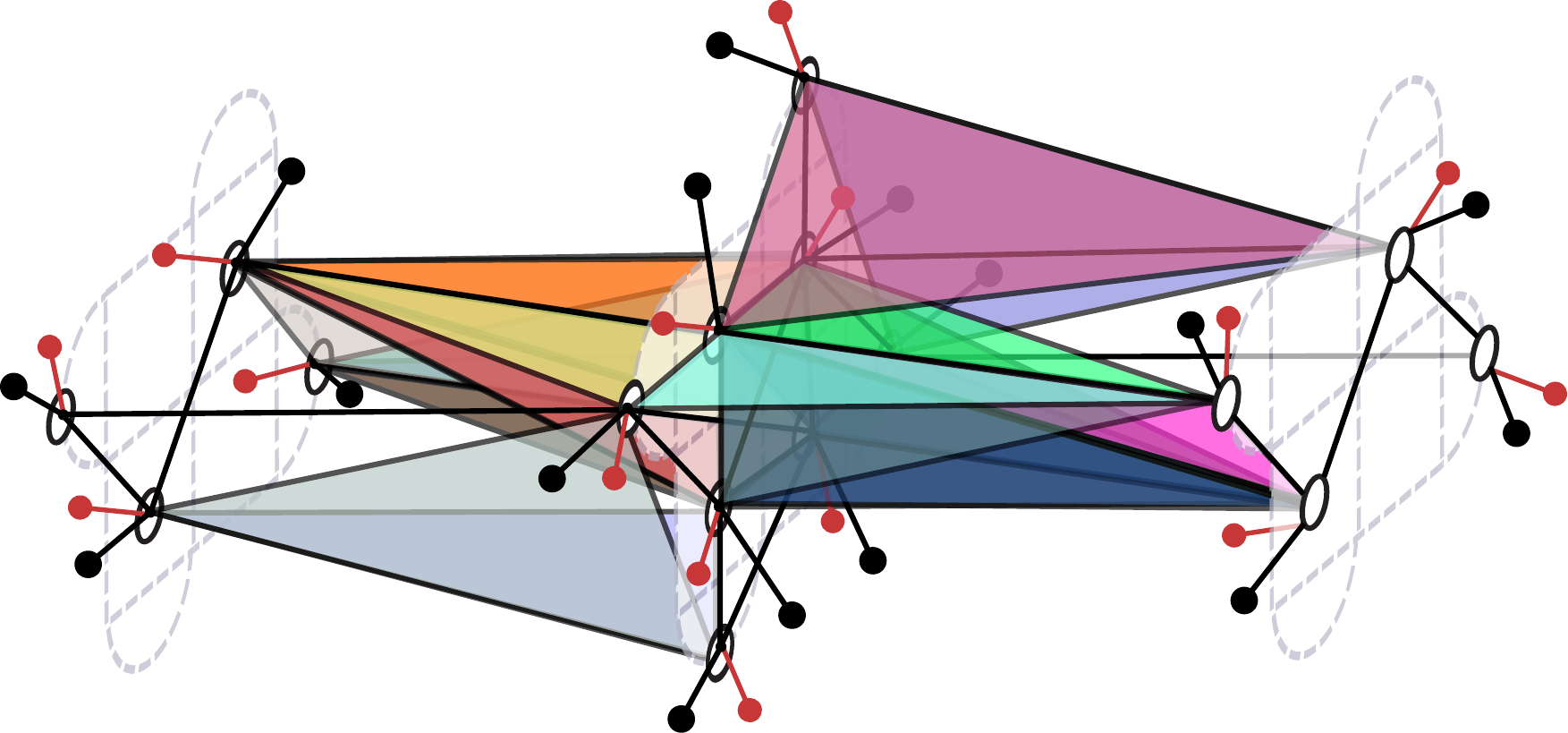}
    \caption{Tensor network representation of the detector error model for a surface code, constructed in the detector picture. Each XOR tensor enforces detector parity, and probability tensors encode the error mechanism weights.}
    \label{fig:s2}
\end{figure}

To address this problem, we directly consider summing over all possible error configurations and setting the probabilities of those configurations that do not satisfy the constraints to zero:
\begin{equation}
    p_{\bm{\theta}}(\mathbf{s}) = \sum_{\mathbf{e}} p(\mathbf{e})\cdot\delta_{\mathbf{s},\mathbf{s}_\mathbf{e}}\;.
    \label{eq:s5}
\end{equation}

Further, we can write both $\mathbf{e}$ and $\mathbf{s}$ from the above equation in terms of their components:
\begin{equation}
\begin{split}
    \sum_{\mathbf{e}} p(\mathbf{e})\cdot\delta_{\mathbf{s},\mathbf{s}_\mathbf{e}} &= \sum_{\{e_i\}} \prod^n_{i=1} \theta_i^{e_i}(1-\theta_i)^{1-e_i}\prod^m_{j=1} s_j\oplus e_{\partial j} \\
    &=\sum_{\{e_i\}} \prod^n_{i=1} P_{e_i}\prod^m_{j=1} X_{e_{\partial j},s_j}
\end{split}\;,
\end{equation}
where $P_{e_i}$ denotes the probability tensor with components $P_0=1-\theta_i$ and $P_1=\theta_i$. By introducing a copy tensor for each variable $e_i$, the probability tensor effectively acts as a weighted copy tensor possessing two non-zero elements, $1-\theta_i$ and $\theta_i$. Additionally, $X_{e_{\partial j},s_j}$ is the XOR tensor as defined in the main text. This class of tensor network representations is known as the detector picture~\cite{PRXQuantum.5.040303}.

Furthermore, in practical implementations, certain XOR tensors exhibit an excessively large degree, causing the tensor network to exceed the memory capacity of a single GPU during initialization. To address this, we utilize the Walsh-Hadamard transform to decompose a high-degree XOR tensor into a central summation index conjugated by Hadamard matrices on each leg~\cite{PRXQuantum.5.040303}:
\begin{equation}
   X_{ijk\cdots} \propto \sum_{\alpha} H_{\alpha i} H_{\alpha j} H_{\alpha k} \cdots\;.\label{eq:s7}
\end{equation}

This factorization significantly reduces memory overhead, as only the decomposed components need to be stored.
\begin{figure}[h]
    \centering
    \includegraphics[width=0.8\linewidth]{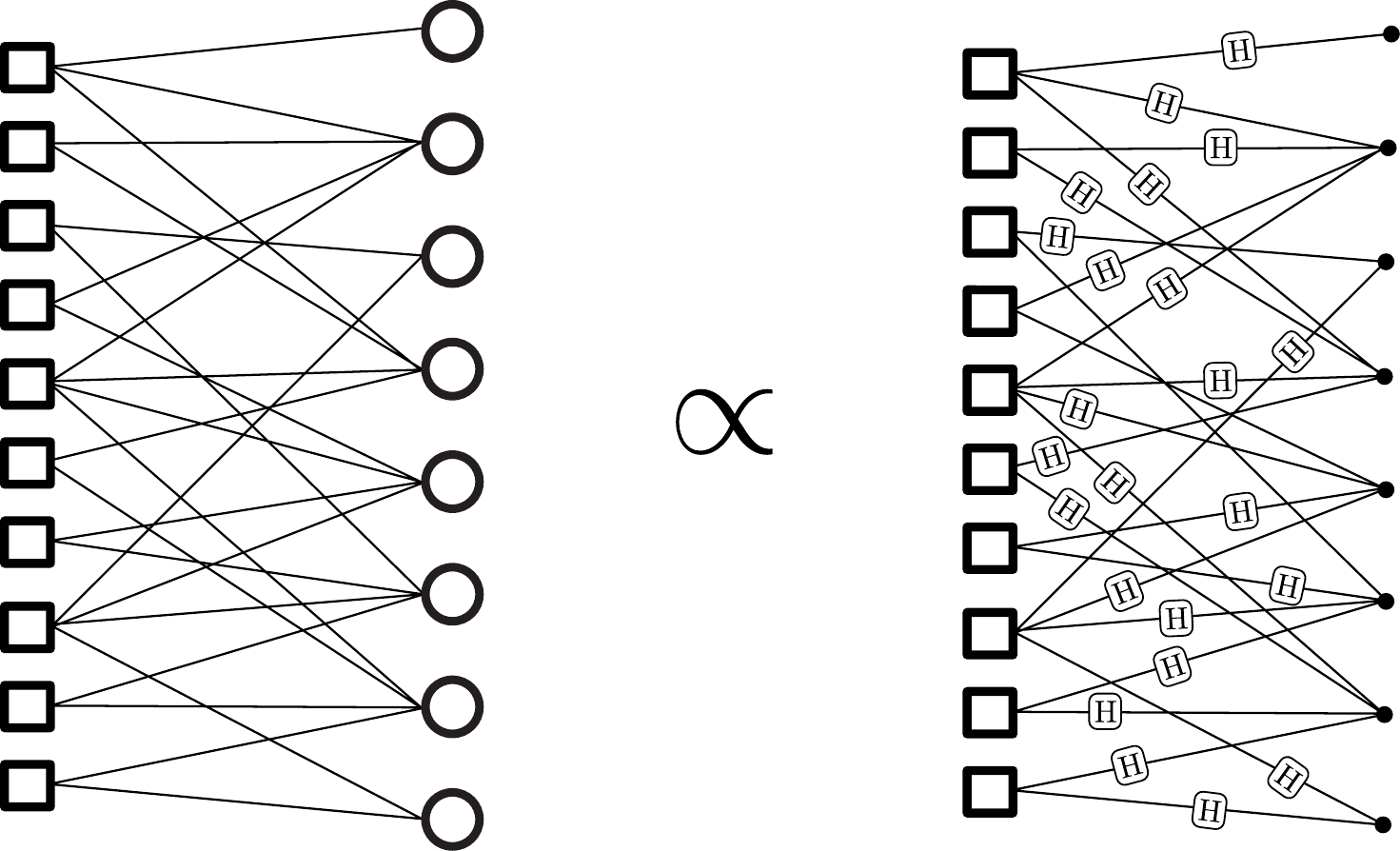}
    \caption{Walsh-Hadamard transform. The left and right sides respectively show the tensor network structures before and after transformation.}
    \label{fig:placeholder}
\end{figure}

Another major advantage of this factorization scheme is that the tensor networks used for both decoding and optimization share an identical structure, differing only in the values of specific tensor elements. For a single logical qubit, maximum likelihood decoding requires comparing $p(l=0)$ and $ p(l=1)$, which is equivalent to computing the sign of the difference, $p(l=0)-p(l=1)$. Calculating this quantity allows us to easily decompose the high-degree XOR tensor associated with the logical operator into a set of vectors. Once these vectors are absorbed into the probability tensors, they merely flip the signs of the tensor elements~\cite{PRXQuantum.5.040303}. We present a brief proof below. Assuming the XOR tensor has been factorized, the contraction of the entire tensor network can be rewritten in the following form:
\begin{equation}
    \sum_i A_iH_{il} = p_l \;,
\end{equation}
where the two elements of $p_l$ represent the probabilities $p(l=0)$ and $p(l=1)$, $H_{il}$ denotes the final Hadamard tensor connected to the logical check vector, and $A_i$ represents the contraction result of the remainder of the TN. Utilizing the unitary property of the Hadamard tensor, we can deduce that:
\begin{equation}
    A_1 \propto p_0-p_1\;.
\end{equation}

Fundamentally, the decoding problem reduces to evaluating $A_1$. Moreover, it is straightforward to see that the index $i$ corresponds directly to the summation index in Eq.~(\ref{eq:s7}). When this index is set to $1$, all connected Hadamard tensors reduce to vectors and can be absorbed into the probability tensors. This absorption process effectively only flips the signs of specific elements (i.e., $\theta\rightarrow-\theta$) within the probability tensors. Consequently, for a given system, we only need to determine an efficient contraction path for a single tensor network structure, which significantly reduces our computational overhead.

\subsection{Subsample Strategy}
\label{sec:subsample}

Computing syndrome likelihoods by TN contraction becomes impractical at larger code distances, as the contraction cost grows exponentially with the number of detectors (Sec.~\ref{sec:complexity}). The subsample strategy resolves this by rewriting the estimation problem itself rather than by approximating the contraction. The larger code is decomposed spatially into smaller overlapping patches with substantially reduced contraction complexity, all sharing the single physical parameter vector $\bm{\theta}$. We define the mean patchwise negative log-likelihood as a diagnostic
\begin{equation}
\mathcal{L}_{\mathrm{patch}}(\bm{\theta})=-\mathbb{E}_{\mathbf{s}\sim p_{\mathrm{data}}(\mathbf{s})}\left[\frac{1}{N_p}\sum_{\alpha=1}^{N_p}\log p_{\bm{\theta},\alpha}(\mathbf{s}_{\alpha})\right],
\label{eq:patch-nll-general}
\end{equation}
where $N_p$ is the number of patches, $\mathbf{s}_{\alpha}$ denotes the corresponding local syndrome sliced from the same shots, and $p_{\bm{\theta},\alpha}(\mathbf{s}_{\alpha})$ is its likelihood given by the patch-TN contraction.

Each patch distribution is the marginal of the full DEM syndrome distribution:
\begin{equation}
p_{\bm{\theta},\alpha}(\mathbf{s}_\alpha)
=\sum_{\mathbf{s}:\,\mathbf{s}|_{D_\alpha}=\mathbf{s}_\alpha}
p_{\bm{\theta}}(\mathbf{s}).
\label{eq:patch-marginal}
\end{equation}
Retaining the restrictions of all mechanisms intersecting the patch makes this
forward marginal exact within the assumed DEM, even for boundary-crossing
mechanisms. The product of these marginal likelihoods is a composite likelihood;
it is not the full-code joint likelihood and need not define a normalized joint
distribution over the full syndrome. Overlapping patches are correlated, but
their overlap does not invalidate the composite loss. It does mean that they
cannot be treated as independent data replicates.

The unmasked population loss has the decomposition
\begin{equation}
\begin{split}
\mathcal{L}_{\mathrm{patch}}(\bm{\theta})
&= C+\frac{1}{N_p}\sum_{\alpha=1}^{N_p}
D_{\mathrm{KL}}\!\left(
p_{\mathrm{data},\alpha}\,\|\,p_{\bm{\theta},\alpha}\right),\\
C&=\frac{1}{N_p}\sum_{\alpha=1}^{N_p}H(p_{\mathrm{data},\alpha}),
\end{split}
\label{eq:patch-kl}
\end{equation}
where $p_{\mathrm{data},\alpha}$ is the data marginal on patch $\alpha$ and $C$
is parameter-independent. If the data are generated by the assumed DEM at
$\bm{\theta}^{*}$, all patch KL divergences vanish there, so
$\bm{\theta}^{*}$ is a global minimizer of this population loss. Uniqueness
requires that the collection of patch marginals jointly identify the parameters:
agreement with every patch marginal must imply agreement of the parameter vector
within the admissible domain. Local regularity instead requires the stacked
Jacobians of the patch marginal distributions to have full column rank at the
true parameters. Neither property follows from hyperedge coverage alone.
Moreover, this population-loss argument does not establish convergence of the
masked update below or parameter recovery from finite experimental data.

\begin{figure}[htbp]
    \centering
    \includegraphics[width=0.5\linewidth]{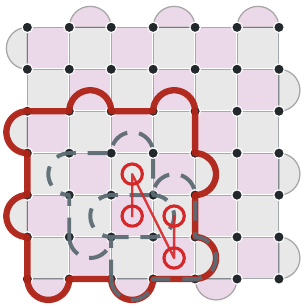}
    \caption{A single error mechanism of the $d=7$ DEM (red rings: four detectors spanning two detector layers) lies entirely inside this $d=5$ patch (red boundary). Grey dashed contours show two placements of a $d=3$ patch, each covering three of the four detectors while leaving a different detector outside. Indeed, no placement of a $d=3$ patch can cover this hyperedge.}
    \label{fig:patch_size}
\end{figure}

At the code level, we divide the code into overlapping patches. Taking the $d=7$ surface code as an example, the code is divided into five overlapping $d=5$ patches ($N_p=5$), as shown in Fig.~\ref{fig:1}(b) (lower): four corner patches and one central patch, placed symmetrically about the lattice center. At the DEM level, each error mechanism is a hyperedge over detectors, and it enters the sub-DEM of a patch whenever it triggers at least one detector of that patch. The patch size is chosen as the smallest one guaranteeing full coverage: every mechanism of the larger code's DEM is fully contained in at least one $d=5$ patch, while $d=3$ patches are insufficient. One such example is illustrated in Fig.~\ref{fig:patch_size}. This coverage ensures that each fitted mechanism is fully represented in at least one patch and can receive an unmasked contribution. It is a geometric requirement of the masked scheme, rather than a proof that the patch marginals identify all parameters. Applying the construction to larger codes requires verifying both coverage and the information retained by the chosen patches.


An error mechanism crossing a patch boundary appears partially in several
patches. We retain these partial occurrences in the forward evaluation, since
they are needed to obtain the exact marginal likelihood
$p_{\bm{\theta},\alpha}(\mathbf{s}_{\alpha})$. In the Willow $d=7$ experiments,
we mask their backward contributions: a parameter receives a contribution from
a patch only if its complete detector support lies inside that patch. Let
$D_i$ be the detector support of mechanism $i$, let $D_\alpha$ be the detectors
in patch $\alpha$, and define
\begin{equation}
M_{i\alpha}=\mathbf{1}\{D_i\subseteq D_\alpha\}.
\end{equation}
For a mini-batch of $B$ shots, the masked negative-log-likelihood gradient
contributions are summed over patches and divided by the total number of
patches $N_p$ (five in the Willow experiments):
\begin{equation}
\begin{split}
g_i^{\mathrm{mask}}(\bm{\theta})
&=-\frac{1}{BN_p}\sum_{k=1}^{B}\sum_{\alpha=1}^{N_p}
M_{i\alpha}\,
\frac{\partial}{\partial\theta_i}
\log p_{\bm{\theta},\alpha}(\mathbf{s}^{(k)}_\alpha).
\end{split}
\label{eq:masked-patch-gradient}
\end{equation}
The optimizer uses these aggregated contributions to update the shared parameter
vector. The mask changes only the backward contribution; the forward likelihood
continues to depend on the current values of boundary-crossing mechanisms.

Without masking, this average would be the mini-batch gradient of
$\mathcal{L}_{\mathrm{patch}}$. With parameter-dependent masks,
$\mathbf{g}^{\mathrm{mask}}$ is generally not that gradient, nor do we establish
that it is the gradient of another scalar objective. We therefore describe the
Willow procedure as a masked likelihood-gradient update, rather than as ordinary
gradient descent minimizing Eq.~\eqref{eq:patch-nll-general}. A decrease in the
monitored patchwise loss is an empirical observation, not a guaranteed property
of this update. Exact patch likelihoods should likewise be distinguished from
exact full-code maximum-likelihood estimation.

Masking is an empirical choice motivated by parameter recovery on finite data.
The unmasked contributions are valid derivatives of the patch marginals; we do
not characterize them as incorrect or contaminated gradients. The convergence
analysis below concerns the unmasked full-network likelihood and does not provide
a convergence guarantee for Eq.~\eqref{eq:masked-patch-gradient}.

\begin{figure}[htbp]
    \centering
    \includegraphics[width=0.8\linewidth]{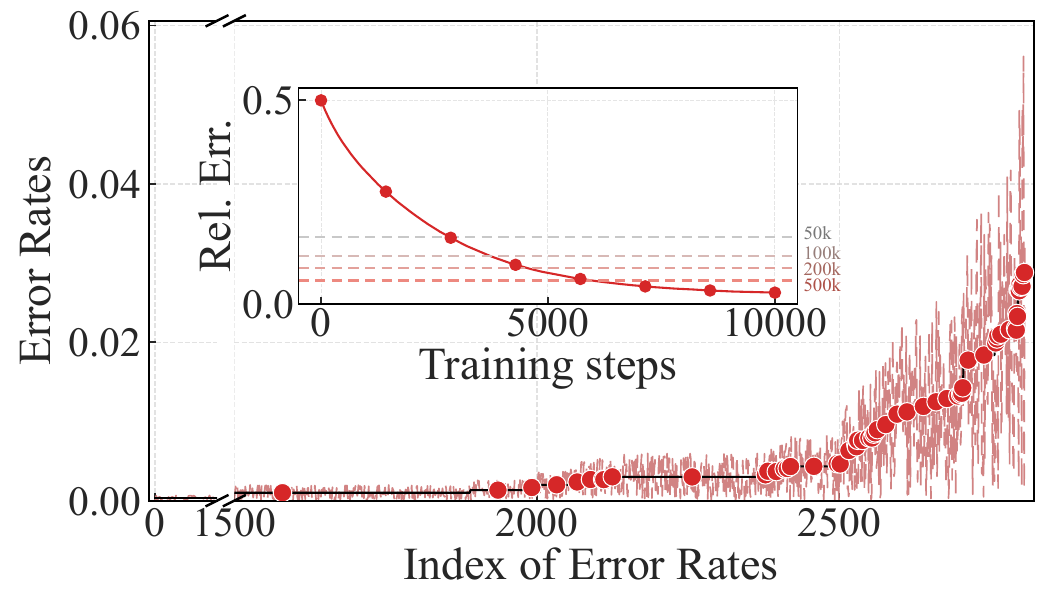}
    \caption{DEM parameter recovery for a simulated $d=7$, $r=4$ surface code $Z$-basis memory experiment with $\epsilon_p=0.005$. Symbols and markers are consistent with Fig.~\ref{fig:sim}. The axis break omits indices 50--1500. The inset shows the mean relative error over all 2807 parameters versus training steps. At each training step, 1024 fresh syndrome samples are drawn from the ground-truth DEM for training; the dashed lines mark the final relative errors obtained using fixed datasets of 50k, 100k, 200k, and 500k samples.}
    \label{fig:d7_recovery}
\end{figure}
Following the simulation protocol of Sec.~\ref{sec:statmech}, we validate the subsample strategy on simulated $d=7$, $r=4$ surface-code data generated with Stim for a $Z$-basis memory experiment at $\epsilon_p=0.005$. As shown in Fig.~\ref{fig:d7_recovery}, with 1024 fresh syndrome samples drawn from the ground-truth DEM at every training step, the error probabilities are recovered close to the ground truth from perturbed initial priors once training converges. We also test the training with fixed datasets, where the Monte Carlo gradient noise becomes significant when data are limited. This effect is more pronounced at larger code distances. Repeating the experiment with fixed training sets of 50k, 100k, 200k, and 500k samples, the relative error of the converged DEM decreases monotonically as the dataset grows, as marked by the dashed lines in the inset of Fig.~\ref{fig:d7_recovery}.

\begin{figure*}[t!]
    \includegraphics[width=0.93\linewidth]{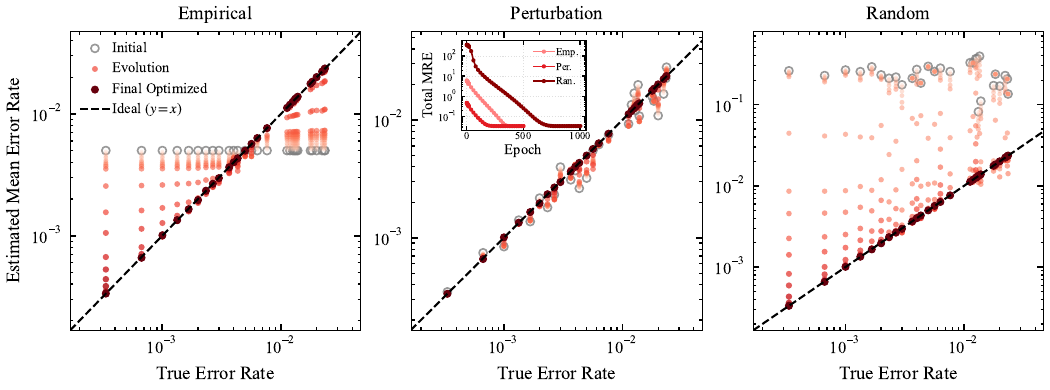}
    \caption{\textbf{Simulation results.} Estimated versus true error rates for a $d=3$, $r=5$ surface code (base depolarizing error rate $0.005$) with 1,000,000 shots. Error mechanisms sharing the same true error rate are grouped, and the mean of their estimated rates (``Estimated Mean Error Rate'') is plotted. We compared three distinct initial conditions: empirical assignment (Emp., uniform $0.005$), local perturbation (Per.), and completely random initialization (Ran.). The color gradient illustrates the evolutionary trajectory of the parameters over successive training epochs. Inset: mean relative error over all parameters (``Total MRE''), $\frac{1}{n}\sum_i|\theta_i-\theta_i^{*}|/\theta_i^{*}$ with $\theta_i^{*}$ the true rates, versus epoch for the three initializations.}\label{fig:init}
\end{figure*}

\begin{figure*}[t!]
    \includegraphics[width=0.93\linewidth]{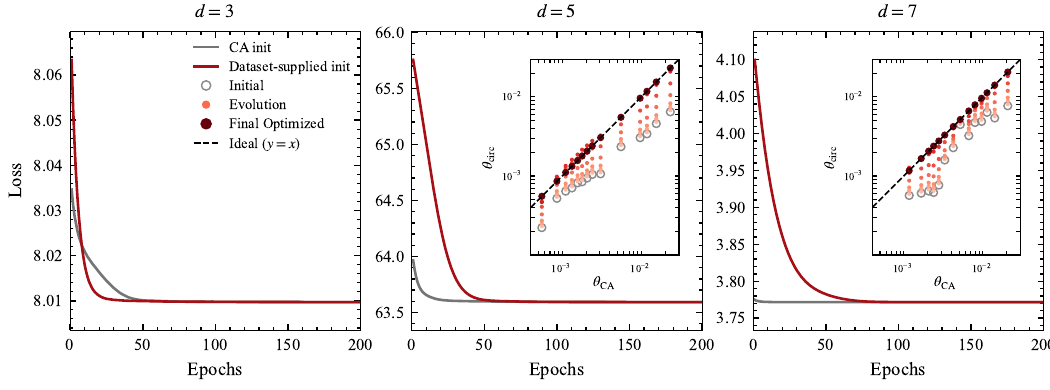}
    \caption{\textbf{Convergence consistency on experimental data.} On the $d{=}3$ ~\cite{dataset_old}, $d{=}5$ ~\cite{dataset} and $d{=}7$ ~\cite{GoogleQuantumAI2024WillowData} experimental surface-code datasets, the estimation procedure is run twice on the same detection events, from two initial DEMs: a correlation-based prior (CA init, gray; this is the SE prior at $d{=}7$) and the dataset-supplied prior (Dataset-supplied init, dark red; the calibrated noisy circuit for $d{=}3$, the uninformative \texttt{simple.dem} for $d{=}5$, and the SI1000 noisy circuit for $d{=}7$). Main panels: loss versus epoch; both initializations converge to the same loss. For $d{=}3$ and $d{=}5$ the loss is the negative log-likelihood per shot of the full DEM, $-\langle\ln p_{\bm{\theta}}(\mathbf{s})\rangle$, in nats; for $d{=}7$ it is the monitored patchwise loss of Eq.~(\ref{eq:patch-nll-general}) up to normalization, evaluated during masked-gradient training, so the absolute scales of the panels are not comparable. Insets (shown for $d{=}5$ and $d{=}7$): error rates optimized from the dataset-supplied initialization, $\theta_{\mathrm{circ}}$, versus those from the CA initialization, $\theta_{\mathrm{CA}}$; symbol conventions follow Fig.~\ref{fig:init}, and the dashed line is $y=x$. The two converged DEMs show agreement in the cases tested. This supports empirical convergence consistency for these initializations, rather than independence from arbitrary initializations on experimental data.}\label{fig:converge_exp}
\end{figure*}

\subsection{Convergence Consistency}

The analytical argument in this subsection applies to the unmasked full-network
negative log-likelihood under the assumptions stated below. It does not establish
local strong convexity or convergence for the masked update used on Willow,
Eq.~\eqref{eq:masked-patch-gradient}. The $d=7$ initialization comparison is
separate empirical evidence for that update.

Let $\mathbf{s}$ denote a syndrome outcome and let $p_{\text{data}}(\mathbf{s})$
be the corresponding data distribution. The model assigns to each syndrome a
probability $p_{\bm{\theta}}(\mathbf{s})$, computed exactly from the DEM priors
$\bm{\theta}\in\mathbb{R}^{n}$ via the \textit{Planar} or TN summation, where $n$
is the number of error mechanisms. The training objective is the negative
log-likelihood
\begin{equation}
\mathcal{L}(\bm{\theta}) = -\mathbb{E}_{\mathbf{s} \sim p_{\text{data}}(\mathbf{s})} \left[ \log p_{\bm{\theta}}(\mathbf{s}) \right].
\label{eq:loss-sm}
\end{equation}
Since
$\mathcal{L}(\bm{\theta}) = H(p_{\text{data}}) +
D_{\mathrm{KL}}\!\left(p_{\text{data}}\,\|\,p_{\bm{\theta}}\right)$
with $H(p_{\text{data}})$ the $\bm{\theta}$-independent entropy of the data
distribution, minimizing \eqref{eq:loss-sm} is exactly equivalent to minimizing the
Kullback--Leibler divergence between $p_{\text{data}}$ and the model prediction
$p_{\bm{\theta}}$. Let $\bm{\theta}^{*}$ be the true underlying physical
parameter vector. We make the following standing assumptions:
(A1)~every fitted error mechanism has a nonzero syndrome signature, and no two
fitted mechanisms induce the same syndrome signature; mechanisms invisible to
the syndromes are excluded, and mechanisms with identical signatures cannot be
identified separately from syndrome data alone;
(A2)~each physical mechanism probability satisfies $\theta_{i}\in[0,\,0.5]$, and
$\bm{\theta}^{*}$ lies in the interior of this domain;
(A3)~the data distribution is realizable by the model at the true parameters,
$p_{\text{data}}(\mathbf{s}) = p_{\bm{\theta}^{*}}(\mathbf{s})$ for all
$\mathbf{s}$ (in practice, $p_{\text{data}}$ is estimated from a finite number
of shots and converges to $p_{\bm{\theta}^{*}}$ as the sample size grows).

Because the exact summation provides a fully differentiable map
$\bm{\theta}\mapsto p_{\bm{\theta}}(\mathbf{s})$, the gradient of
\eqref{eq:loss-sm} takes the score-function form
\begin{equation}
\frac{\partial\mathcal{L}}{\partial\theta_{i}}
\;=\; -\mathbb{E}_{p_{\text{data}}}\!\left[
\frac{\partial\log p_{\bm{\theta}}(\mathbf{s})}{\partial\theta_{i}}\right].
\label{eq:loss-grad}
\end{equation}
At $\bm{\theta}^{*}$, assumption (A3) gives
$p_{\text{data}} = p_{\bm{\theta}^{*}}$, so that
\begin{equation}
\frac{\partial\mathcal{L}}{\partial\theta_{i}}(\bm{\theta}^{*})
\;=\; -\mathbb{E}_{p_{\bm{\theta}^{*}}}\!\left[
\partial_{i}\log p_{\bm{\theta}}(\mathbf{s})\right]_{\bm{\theta}^{*}}
\;=\; -\partial_{i}\sum_{\mathbf{s}}p_{\bm{\theta}^{*}}(\mathbf{s})
\;=\; 0,
\label{eq:loss-stationary}
\end{equation}
i.e.\ $\bm{\theta}^{*}$ is a stationary point of the loss. Differentiating
\eqref{eq:loss-grad} once more and evaluating at $\bm{\theta}^{*}$ yields the
Hessian
\begin{equation}
\frac{\partial^{2}\mathcal{L}}
{\partial\theta_{i}\partial\theta_{j}}(\bm{\theta}^{*})
\;=\; -\mathbb{E}_{p_{\bm{\theta}^{*}}}\!\left[
\frac{\partial^{2}\log p_{\bm{\theta}}(\mathbf{s})}
{\partial\theta_{i}\partial\theta_{j}}\right]_{\bm{\theta}^{*}}
\;=\; \mathcal{I}_{ij}(\bm{\theta}^{*}),
\label{eq:loss-hess}
\end{equation}
where the last equality uses the classical Fisher-information identity
$$\mathbb{E}_{p_{\bm{\theta}}}[\partial_{i}\partial_{j}\log p_{\bm{\theta}}]
= -\,\mathbb{E}_{p_{\bm{\theta}}}[\partial_{i}\log p_{\bm{\theta}}\,\partial_{j}\log p_{\bm{\theta}}]\;,$$
valid for any normalized parametric family. Equivalently,
\begin{equation}
\begin{split}
&\nabla^{2}\mathcal{L}(\bm{\theta}^{*})
\;=\; G(\bm{\theta}^{*})^{\top}\,W(\bm{\theta}^{*})\,G(\bm{\theta}^{*}),
\qquad\\
&W(\bm{\theta}^{*}) = \mathrm{diag}\!\left(\frac{1}{p_{\bm{\theta}^{*}}(\mathbf{s})}\right)_{\mathbf{s}},
\label{eq:loss-hess-jwj}
\end{split}
\end{equation}
where $G(\bm{\theta}^{*})$ is the Jacobian of the map
$\bm{\theta}\mapsto\big(p_{\bm{\theta}}(\mathbf{s})\big)_{\mathbf{s}}$, with the
outcomes restricted to the reachable syndrome support of the fixed DEM. Outcomes
outside this support have zero probability for all parameters and are omitted.
Since every reachable syndrome has strictly positive probability under (A2),
the diagonal weights $1/p_{\bm{\theta}^{*}}(\mathbf{s})$ are strictly positive,
so $W(\bm{\theta}^{*})\succ 0$. Hence $\nabla^{2}\mathcal{L}(\bm{\theta}^{*})$
is positive definite if and only if $G(\bm{\theta}^{*})$ has full column rank,
which is precisely the local identifiability of the parameterization.
Assumption (A1) guarantees this: if no two error mechanisms induce the same
syndrome signature, the columns of $G(\bm{\theta}^{*})$ are linearly independent
and no direction in parameter space leaves the syndrome distribution invariant;
consequently
\begin{equation}
\mathcal{I}(\bm{\theta}^{*})\;\succ\;0.
\label{eq:fisher-pd}
\end{equation}
Strict positive-definiteness of the Hessian, together with the
twice-differentiability of $p_{\bm{\theta}}$, yields the local strong convexity
of $\mathcal{L}$: there exist $\rho>0$ and $\mu>0$ such that for all
$\bm{\theta}\in B(\bm{\theta}^{*},\rho)$,
\begin{equation}
\mathcal{L}(\bm{\theta})-\mathcal{L}(\bm{\theta}^{*})
\;\ge\;\frac{\mu}{2}\,\|\bm{\theta}-\bm{\theta}^{*}\|^{2}.
\label{eq:lsc}
\end{equation}
By (A2), $\bm{\theta}^{*}$ lies in the interior of the physical domain
$[0,\,0.5]^{n}$, so $B(\bm{\theta}^{*},\rho)$ may be chosen small enough to be
contained in the open cube $(0,\,0.5)^{n}$; any physically admissible
initialization (i.e.\ one respecting $\theta_{i}\in(0,\,0.5)$) and sufficiently
close to $\bm{\theta}^{*}$ therefore lies inside this basin. Within
$B(\bm{\theta}^{*},\rho)$ the loss admits a unique stationary point, namely
$\bm{\theta}^{*}$, and gradient descent with a sufficiently small step size
converges to it monotonically. The convergence point of the optimization is
consequently independent of the initialization in this regime, which accounts
for convergence from sufficiently close initializations, without establishing a
global guarantee for the other initializations tested.

The preceding analysis establishes local strong convexity under its assumptions,
but does not establish global convergence or bound the basin of attraction for
finite-sample training. We therefore examine convergence empirically from three
initializations: empirical assignment, local perturbation, and fully random values.
In Fig.~\ref{fig:init}, the grouped mean estimates approach the $y=x$ line and the
mean relative error over all parameters decreases during training. Agreement of
the grouped means does not by itself establish accurate recovery of every
parameter; these runs provide empirical evidence for the tested initializations. 

We also examined convergence on experimental data, where no ground
truth is available. For the $d=3$ and $d=5$ datasets, we optimized
the dMLE objective twice on the same detection events, starting
from a CA prior and from the prior supplied with
the dataset: the calibrated noisy circuit for $d=3$~\cite{dataset_old}
and the uninformative \texttt{simple.dem} for $d=5$~\cite{dataset}.
As shown in Fig.~\ref{fig:converge_exp}, the two optimizations
converge to the same loss value, and the resulting DEM parameters
are in complete agreement for both datasets. When scaling dMLE
to $d=7$ using the masked subsample update, we observe empirical consistency
between the two runs: starting from the SE prior
and the SI1000~\cite{gidney2021fault} noisy-circuit prior supplied with the dataset~\cite{GoogleQuantumAI2024WillowData},
both initializations yield consistent final DEM estimates and monitored patchwise
loss values. This agreement does not establish a minimum of the patchwise loss
or a general convergence guarantee for the masked update. Although we cannot prove this convergence consistency rigorously, across the experimental datasets we have tested, the two most common empirical initializations, namely the circuit-noise prior supplied with the dataset and a correlation-based prior (CA at $d=3$ and $d=5$, SE at $d=7$), both lead to the same converged result. This indicates that our method, in both its exact and its approximate subsampled form, can start from the most readily available empirical distribution and optimize from there.

\begin{figure}[htb]
    \centering
    \includegraphics[width=0.98\linewidth]{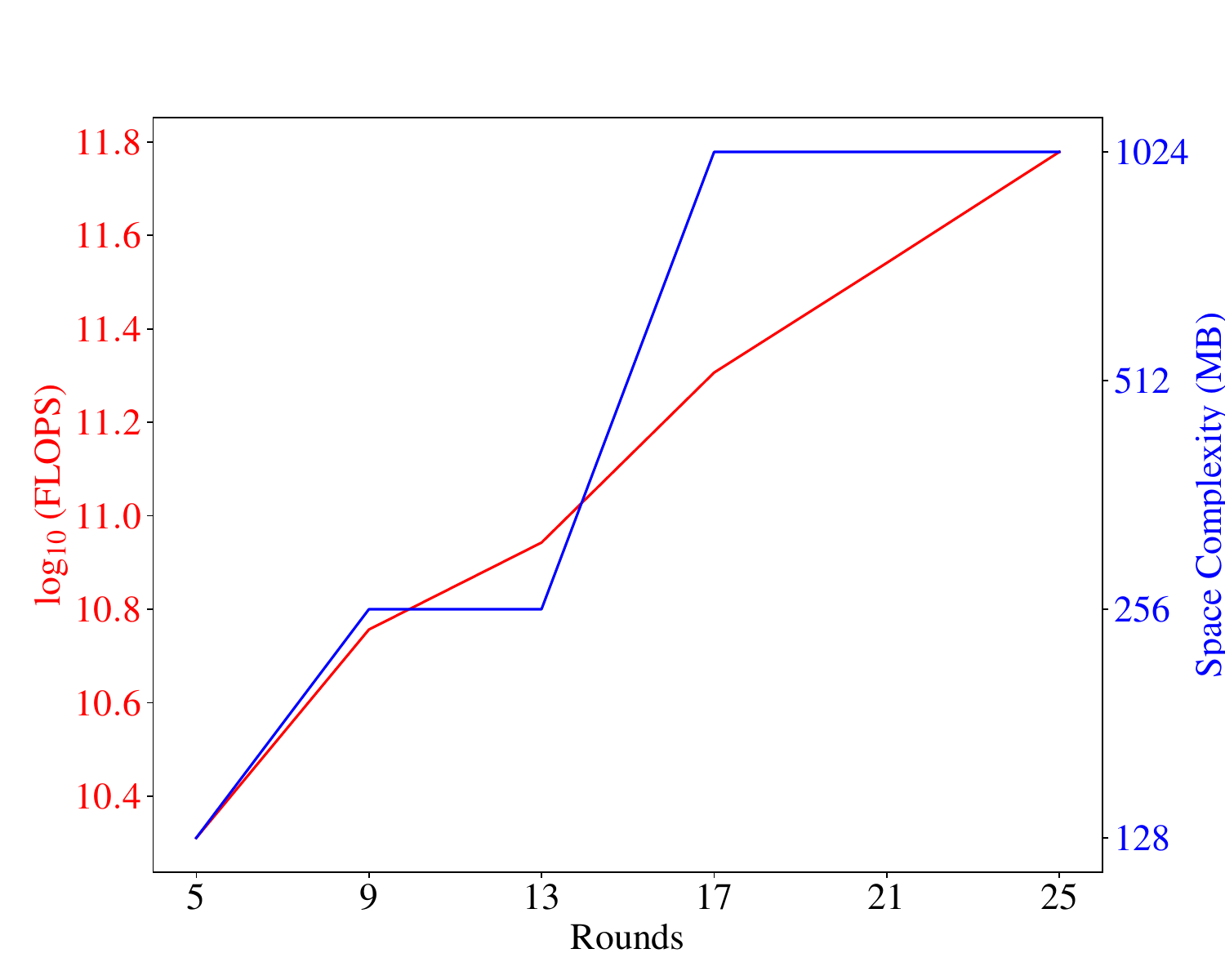}
    \caption{Contraction complexity of the tensor network for maximum likelihood estimation on the $d=5$ rotated surface code as a function of the number of QEC rounds. Left axis: number of floating-point operations (FLOPs) of one contraction; right axis: size of the largest intermediate tensor.}
    \label{fig:tn_complexity}
\end{figure}

\section{Complexity Analysis}
\label{sec:complexity}
By applying the Walsh-Hadamard transformation, we decompose all XOR tensors into Hadamard matrices and copy tensors, with the latter represented using hyper-indices. 
This leaves only two types of components in the tensor network: one-dimensional error probability vectors and Hadamard matrices, dramatically simplifying the network structure underlying our differentiable maximum likelihood estimation method. 
We then employ the tensor network contraction path finder from our previous work~\cite{pan2022Simulation, pan2022Solving, pan2024Efficient}, specifically the OMEinsumContractionOrders.jl~\cite{omeinsum} implementation.
The path finder uses simulated annealing over contraction trees to minimize a cost function combining time complexity, memory cost, and memory access cost, thereby obtaining an efficient contraction order~\cite{kalachev2022multitensorcontractionxebverification}.
The resulting contraction complexity for the $d=5$ rotated surface code with rounds ranging from $r=5$ to $r=25$ is plotted in Fig.~\ref{fig:tn_complexity}.

Our computational strategy uses an optimized contraction order for the
Walsh--Hadamard-transformed network without tensor compression. This differs
from the boundary-MPS approach discussed in previous surface-code decoding
work~\cite{google2023suppressing,shutty2024Efficient}; a direct timing comparison
would additionally require matching the network, numerical precision, decoding
task, and hardware. The forward-contraction estimates below quantify the
tractability of our representation and contraction order.

The total contraction cost for $d{=}5$, $r{=}25$ is $6.3\times 10^{11}$ floating-point operations (FLOPs), and the largest intermediate tensor occupies approximately $1$~GB of memory.
On a high-end NVIDIA H200 GPU, which delivers around $3\times 10^{13}$ floating-point operations per second with $141$~GB of memory, we can comfortably batch-contract approximately $32$ syndromes simultaneously, completing each batch in seconds.
For training, the same tensor operations are differentiated in reverse mode
to obtain derivatives with respect to all DEM probabilities. A contraction order
found for a fixed topology is reused for the different syndromes and parameter
updates, enabling batched forward and backward computation. The FLOP and
intermediate-tensor estimates above describe the forward contraction only;
training additionally incurs backward computation and storage for intermediate
values. We do not equate these estimates with measured end-to-end training costs.

Direct evaluation of the syndrome likelihood becomes prohibitively
expensive for larger surface codes. For the full $d=7$, $r=10$ tensor
network, the best contraction order we found requires approximately
$8.9\times10^{18}$ FLOPs, and its largest intermediate tensor occupies
approximately $4.5\times10^{6}$~GB of memory. Our subsampling strategy instead replaces this contraction with five
$d=5$, $r=10$ patch contractions, whose individual costs are shown in
Fig.~\ref{fig:tn_complexity}. Together, the five patch contractions cost
approximately $3.2\times10^{11}$ FLOPs, while the largest intermediate
tensor in each contraction occupies approximately $0.25$~GB; both values
are more than seven orders of magnitude below the corresponding estimates
for the full-network contraction order above.

For larger code distances, the dominant computational cost of the
patchwise scheme scales only with the number of patches,
$N_p\sim5^{(d-5)/2}$, corresponding to 5, 25, and 125 patches for
$d=7$, $d=9$, and $d=11$, respectively. Although the number of patches
grows exponentially, the cost of each patch does not grow with the
target code distance, and the total cost remains far below that of
contracting the full tensor network directly. Moreover, the patch
contractions and gradient calculations are independent and can be
distributed across multiple GPUs before the gradients are aggregated
to update the DEM parameters. In the ideal parallel limit, assigning
one GPU to each patch makes the wall-clock training time comparable
to that required to train a single $d=5$ patch.

\section{Implementation Details}
\begin{table*}[htbp]
  \centering
  \caption{Hyperparameters for the numerical experiments.}
  \label{tab:hyperparams}
  \begin{tabular}{lcccccc}
    \toprule
    & \textbf{\makecell{Rep.\\sim.}} & \textbf{\makecell{Rep.\\BAQIS}} & \textbf{\makecell{SC\\sim.}} & \textbf{\makecell{SC Google\\$d=3$~\cite{dataset_old}}} & \textbf{\makecell{SC Google\\$d=5$~\cite{dataset}}} & \textbf{\makecell{SC Google\\$d=7$~\cite{GoogleQuantumAI2024WillowData}}} \\
    \midrule
    Error rate    & 0.001     & --        & 0.001, 0.005 & --               & --                  & -- \\
    Num. shots    & 1,000,000 & 1,000,000 & 1,000,000 & 96 $\times$ 50,000   & 24 $\times$ 75,000   & 2 $\times$ 50,000 \\
    Epochs        & 500       & 500       & 1,000     & 200                 & 200                 & 100 \\
    Batch size    & 10,000    & 10,000    & 10,000    & 10,000              & 15,000              & 1,024 \\
    Learning rate & 0.001     & 0.001     & 0.001     & 0.01                & 0.01                & 0.01 \\
    \bottomrule
  \end{tabular}
\end{table*}
The training hyperparameters for the different numerical experiments are summarized in Table~\ref{tab:hyperparams}.
Its simulation columns cover the simulated data used in Figs.~\ref{fig:sim} and \ref{fig:init}; the surface-code simulations use $\epsilon_p=0.001$ in Fig.~\ref{fig:sim} and $\epsilon_p=0.005$ in Fig.~\ref{fig:init}. Two simulations use a different setup and are therefore not included in the table: the exact-likelihood benchmark of Fig.~\ref{fig:toy}, which involves no sampling and is run for 2000 epochs, and the subsample test of Fig.~\ref{fig:d7_recovery}, whose details are specified in its caption.

The Google $d=5$ Sycamore dataset used in the RL study~\cite{dataset} consists of 53 subsets, each with 75,000 shots per round setting. Due to computational limitations, processing the entire corpus was not feasible. Instead, our study utilizes a partial sample of 24 subsets, including 10 randomly chosen indices $(0, 2, 10, 12, 14, 16, 24, 32, 46, 49)$ and 14 additionally selected indices $(3$--$9, 11, 13, 15, 17$--$20)$. For each selected subset, we train a separate DEM using only its $r=5$ data and transfer the learned parameters to the other available round settings for decoding; the decoding results reported in the main text are evaluated on these other round settings. We opted not to train a single, universal set of parameters using the combined data because of the substantial heterogeneity across the subsets. As evidenced by the large error bars reported in the original reference~\cite{RL-PhysRevLett.133.150603}, the underlying error characteristics and distributions vary considerably from one subset to another. Consequently, a single model trained on the combined data may fail to capture the distinct error profiles of individual configurations. By training a separate model for each subset, we allow the optimization to adapt to its corresponding experimental noise profile. Our optimization yields consistent improvements for both the TN and Tesseract decoders across all evaluated subsets. These results demonstrate the performance of our method on the selected subsets, while comprehensive validation over the full dataset is left for future work.

In addition, we evaluated our approach on the earlier Google $d=3$ Sycamore dataset~\cite{dataset_old} to assess whether the dMLE prior enables TN decoding to approach the performance reported for the state-of-the-art AlphaQubit decoder~\cite{bausch2024learning}. This dataset contains measurements in both the $X$ and $Z$ bases, with four center locations, $(3,5)$, $(5,3)$, $(5,7)$, and $(7,5)$, and 12 measurement-round settings ranging from $r=3$ to $r=25$ for each basis. Because each subset has a distinct DEM structure and lacks translational symmetry across rounds, we trained a separate dMLE model for each individual subset. Using each resulting dMLE prior, we performed TN decoding on the experimental data. The resulting logical error rate per round is comparable to that reported for AlphaQubit, whose training uses additional experimental information, including leakage data, whereas our DEM estimation uses standard syndrome data alone. We emphasize that, because the DEM structures of the different centers and rounds in the Google $d=3$ dataset share no common repetitive pattern, we used all the syndrome data as the training set. Although training does not use the logical labels, the decoding results are evaluated on the same syndrome data used for DEM estimation and should therefore be interpreted as performance after calibration on that data, rather than as an independent test of generalization. We optimize a generative model of the syndrome distribution, rather than a supervised classifier of logical labels.

\begin{figure}[htbp]
    \centering
    \includegraphics[width=\linewidth]{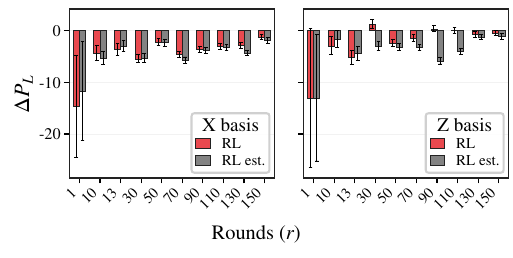}
    \caption{Logical error probability comparison for the $d=7$ surface-code memory experiments on Google Willow, obtained with our reproduction of the method of Ref.~\cite{takou2026logical}. All priors are evaluated with the MWPM decoder (PyMatching). Following Ref.~\cite{takou2026logical}, $\Delta P_L=(P_L-P_L^{\mathrm{SI1000}})/P_L^{\mathrm{SI1000}}\times100\%$ is the percent change of the logical error probability $P_L$ relative to the SI1000 prior, so that negative values correspond to a lower logical error probability. Red bars show the RL prior and gray bars (``RL est.'') show the syndrome-estimated (SE) prior, i.e.\ the DEM probabilities estimated from the syndrome data with the method of Ref.~\cite{takou2026logical} on the error-mechanism support of the RL DEM (their ``est RL'' variant, which differs from the RL prior only in the probabilities).}
    \label{fig:takou}
\end{figure}
Finally, we evaluated our approach on the Google Willow dataset~\cite{GoogleQuantumAI2024WillowData}. For the $X$- and $Z$-basis memory experiments, we independently trained one DEM for each basis using the corresponding $d=7$, $r=10$ data. Each DEM was trained using five overlapping $d=5$ patches and the masked update of Eq.~(\ref{eq:masked-patch-gradient}), with retained gradient contributions summed over patches and divided by the total patch count $N_p=5$. We monitored the mean patchwise loss of Eq.~(\ref{eq:patch-nll-general}); the masked update is generally not its full gradient. We then used the resulting dMLE priors for decoding the data from $r=13$ to $r=250$.
For the $d=7$ surface-code comparison in the main text, we construct the SE prior using the method of Takou \textit{et al.}~\cite{takou2026logical}, the multi-detector generalization of the pairwise correlation analysis of Ref.~\cite{google2021exponential}: the relevant detector correlators are averaged over the same shots and the resulting relations are inverted hierarchically, from four-detector supports down to graphlike and boundary ones, to give the DEM probabilities. To validate our reproduction of their method, Fig.~\ref{fig:takou} shows the resulting change of the MWPM logical error probability relative to the SI1000 prior, in the same form as Fig.~2 of Ref.~\cite{takou2026logical}; our values closely match those reported there. For presentation, we show rounds up to $r=150$.

The Tesseract decoder was configured with \texttt{det\_beam}=20, \texttt{beam\_climbing}=\texttt{True}, \texttt{no\_revisit\_dets}=\texttt{True}, \texttt{pqlimit}=200000, and \texttt{det\_penalty}=0.1. 

The MWPF decoder used \texttt{SinterMWPFDecoder} with
  \texttt{decoder\_type} set to \texttt{SolverSerialJointSingleHair}, \texttt{cluster\_node\_limit} set to \texttt{None}, and \texttt{omp\_thread\_count}=1. 

The RelayBP decoder used \texttt{RelayDecoderF32} with \(\gamma_0=0.1\), \texttt{pre\_iter}=80, \texttt{num\_sets}=60,
  \texttt{set\_max\_iter}=60, \texttt{gamma\_dist\_interval}=(-0.1,0.8), and \texttt{stop\_nconv}=1; the tuning stage used seed 20260908, whereas the production batch runs used seed 0. 
  
For CM, PyMatching was run with \texttt{enable\_correlations}=\texttt{True} during both matching construction and batch decoding; all other PyMatching options used their package defaults.

\section{More Analysis}

\begin{figure}[htbp]
    \centering
    \includegraphics[width=0.9\linewidth]{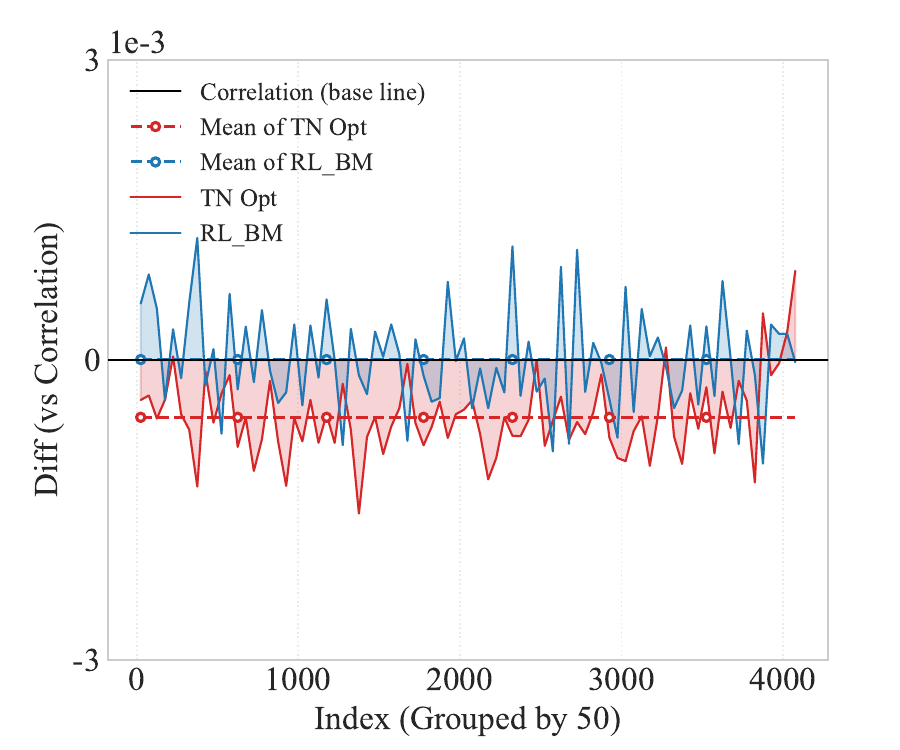}
    \caption{Differences of the dMLE prior (red, labelled ``TN Opt'') and of the RL prior trained with BM (blue, labelled ``RL\_BM'') from the CA prior, for sample$\_00$ ($d=5$, $r=9$).}
    \label{fig:dem_comparison}
\end{figure}

To further elucidate the distinct characteristics of the DEMs, Fig.~\ref{fig:dem_comparison} presents a comparative analysis of the DEMs. 
We specifically contrast the dMLE prior and the RL prior trained with BM against the CA prior.
A detailed examination of the specific instance sample$\_00$ ($d=5$, $r=9$) reveals a critical distinction in optimization strategies.
As shown, the mean error rate of the RL prior remains quantitatively proximal to that of the CA prior.

Conversely, the dMLE prior exhibits a substantial deviation.
This suggests that the RL approach functions primarily as a local fine-tuning mechanism constrained by the CA initialization, rather than learning a fundamentally distinct error model.
This heavy reliance on the CA prior offers a plausible explanation for the RL method's limited generalization capability across different decoders, as observed in our broader experiments.



\end{document}